# Coupling between pore formation and phase separation in charged lipid membranes

Hiroki Himeno[1,2], Hiroaki Ito[3], Yuji Higuchi[4], Tsutomu Hamada[1], Naofumi Shimokawa[1*], Masahiro Takagi[1]

1. School of Materials Science, Japan Advanced Institute of Science and Technology
2. Health Research Institute, National Institute of Advanced Industrial Science and Technology (AIST)
3. Department of Physics, Graduate School of Science, Kyoto University
4. Institute for Materials Research, Tohoku University

*Corresponding author
 Email address: nshimo@jaist.ac.jp

**Abstract**

We investigated the effect of charge on the membrane morphology of giant unilamellar vesicles (GUVs) composed of various mixtures containing charged lipids. We observed the membrane morphologies by fluorescent and confocal laser microscopy in lipid mixtures consisting of a neutral unsaturated lipid [dioleoylphosphatidylcholine (DOPC)], a neutral saturated lipid [dipalmitoylphosphatidylcholine (DPPC)], a charged unsaturated lipid [dioleoylphosphatidylglycerol (DOPG$^{(-)}$)], a charged saturated lipid [dipalmitoylphosphatidylglycerol (DPPG$^{(-)}$)], and cholesterol (Chol). In binary mixtures of neutral DOPC/DPPC and charged DOPC/DPPG$^{(-)}$, spherical vesicles were formed. On the other hand, pore formation was often observed with GUVs consisting of DOPG$^{(-)}$ and DPPC. In a DPPC/DPPG$^{(-)}$/Chol ternary mixture, pore-formed vesicles were also frequently observed. The percentage of pore-formed vesicles increased with the DPPG$^{(-)}$ concentration. Moreover, when the head group charges of charged lipids were screened by the addition of salt, pore-formed vesicles were suppressed in both the binary and ternary charged lipid mixtures. We discuss the mechanisms of pore formation in charged lipid mixtures and the relationship between phase separation and the membrane morphology. Finally, we reproduce the results seen in experimental systems by using coarse-grained molecular dynamics simulations.

## Ⅰ. INTRODUCTION

The basic structure of a biomembrane is a lipid bilayer that is composed of various types of phospholipids. Biomembranes not only separate the inner and outer environments of living cells, but also play a role in a wide range of life-related phenomena through dynamic structural changes. In biomembranes, the components are not uniformly dispersed, and it is believed that such compositional heterogeneity emerges spontaneously. This heterogeneous structure is known as a "lipid raft" [1–3]. Lipid rafts, which are enriched with saturated lipids, cholesterol, or various membrane proteins, are expected to function as platforms to which proteins are attached during signal transduction and membrane trafficking [4,5]. Synthetic lipid vesicles consisting of several lipid molecules are commonly used as models of biomembranes to investigate the physicochemical properties of lipid membranes. In particular, ternary lipid mixtures consisting of a saturated lipid, unsaturated lipid, and cholesterol exhibit phase separation between the saturated lipid and the cholesterol-rich phase (the liquid-ordered ($L_o$) phase) and the unsaturated lipid-rich phase (the liquid-disordered ($L_d$) phase) [6,7]. The spontaneous domain formation that results from this phase separation has attracted great attention in connection to raft formation in biomembranes.

Most previous studies have investigated the primary physical properties of lipid membranes composed of electrically neutral lipids [6,8]. However, biomembranes also contain negatively charged lipids. For instance, the membranes of prokaryotes such as *Staphylococcus aureus* and *Escherichia coli* contain high concentrations of phosphatidylglycerol ($PG^{(-)}$) [9]. In the case of eukaryotic plasma membranes, some organelles such as lysosomes [9] and mitochondria [10,11] are enriched with charged lipids. When we consider physicochemical properties of lipid membranes, the contribution of charged lipids is an important factor to be taken into account. For example, cardiolipin ($CL^{(-)}$), which is enriched in the inner membrane of mitochondria, is considered to contribute to the formation of invagination structures called cristae [12]. In addition, the surface charge of a membrane can be controlled by cellular ions such as sodium, and calcium ions. Several related studies on giant unilamellar lipid vesicles (GUVs) which focused on charged lipids have investigated the effect of charge on lateral phase separation. In ternary mixtures of an unsaturated lipid, a saturated lipid, and cholesterol, negatively charged unsaturated lipids have been shown to suppress phase separation [13–16], and cytochrome c, which is a positively charged protein, induced the formation of micron-sized domains due to electrostatic interaction between charged unsaturated lipids [17]. In our recent study, we investigated the phase behavior induced by negatively charged lipids in several mixtures [18]. We reported that phase separation is suppressed by a charged unsaturated lipid, whereas it is enhanced by a charged saturated lipid.

In addition to the phase separation, membrane morphological changes are an important physicochemical property of biomembranes. Regulation of the membrane morphology is critical for many cellular processes, such as budding in endo- and exocytosis [19], the fission-fusion sequence of vesicular transport [20–22], and pore formation in autophagy [23,24]. These phenomena have inspired several recent studies on, morphological changes of GUVs induced by various external stimuli. For example, GUVs exhibit various morphological changes when they are subjected to osmotic pressure [25]. In addition, pore formation on GUVs and the resulting shrinkage of GUVs were observed under the addition of surfactants [26,27]. With the use of a photo-sensitive amphiphile (KAON12), we successfully controlled the membrane morphology of GUVs to create various forms such as disc, bowl, cup, and sphere shapes, by photo-switching [28].

Moreover, coupling between phase separation and the membrane morphology has been suggested in biomembranes. Recent studies have shown that lipid rafts play a critical role in membrane dynamics such as autophagy and endocytosis [29–31]. In synthetic lipid vesicle systems, we previously reported that inner-budding as in endocytosis is observed under application of osmotic pressure in phase-separated GUVs [32]. In this system, phase-separated domains budded as small vesicles inside GUVs. In addition, Sakuma *et al*. investigated the membrane morphology in binary mixtures of two lipids with different molecular shapes. In binary mixtures of inverse cone- and cylinder-shaped lipids, adhesion between GUVs was observed below the miscibility temperature [33]. On the other hand, in binary mixtures of cone- and cylinder-shaped lipids, GUVs with a large pore were observed [34]. These membrane morphologies were explained in terms of a local change in composition induced by phase separation. However, it is not well understood how the biomembrane structure is controlled in living cells, since most of the previous studies on the morphology of synthetic lipid vesicles were performed in neutral lipid membrane systems.

In a study that focused on the membrane morphology in charged lipid GUVs, Beales *et al*. reported that cytochrome c induces the collapse of domains enriched with cardiolipin ($CL^{(-)}$), which is a negatively charged unsaturated lipid, in four-component mixtures of an unsaturated lipid (DOPC), a saturated lipid (DPPC), cholesterol, and $CL^{(-)}$ [35]. In addition, cytochrome c causes small pore formation in membranes containing DOPC and $CL^{(-)}$ [36]. Riske *et al*. studied the morphological transition of GUVs under a change in temperature in a negatively charged saturated lipid [dimyristoylphosphatidylglycerol ($DMPG^{(-)}$)] single-component system [37]. Within a temperature range from 25℃ to 28℃, a large number of pores were formed on GUVs consisting of $DMPG^{(-)}$. However, the physico-chemical behaviors of the inherent structure in multicomponent lipid membranes and the morphological transition have arrarently not yet been examined systematically, and thus the fundamental coupling between phase separation and membrane morphology in charged lipid mixtures remains unclear.

In this study, we investigated the effect of charge on the membrane morphology in binary mixtures of an unsaturated lipid and a saturated lipid, and ternary mixtures of a neutral saturated lipid, a charged saturated lipid and cholesterol. In particular, we discuss the relation between phase separation and the change in membrane morphology. We observed the membrane morphology of GUVs by using fluorescence microscopy and confocal laser scanning microscopy. We also explored the effect of salt on the membrane morphology. Furthermore, we examined the stability of the membrane morphology under change in temperature. We discuss the mechanism of membrane morphology in charged lipid GUVs, and the relationship between lateral phase separation and membrane morphology. Finally, we reproduced the results seen in experimental systems using coarse-grained molecular dynamics simulations.

## II. MATERIALS AND METHODS
### A. Materials

A neutral unsaturated lipid [1,2-dioleoyl-*sn*-glycero-3-phosphocholine (DOPC, with chain melting temperature, $T_m$= -20℃)], a neutral saturated lipid [1,2-dipalmitoyl-*sn*-glycero-3-phosphocholine (DPPC, $T_m$ = 41℃)], a negatively charged unsaturated lipid [1,2-dioleoyl-*sn*-glycero-3-phospho- (1'-rac-glycerol) (sodium salt) (DOPG$^{(-)}$, $T_m$=-18℃)], a negatively charged saturated lipid [1,2-dipalmitoyl-*sn*-glycero-3-phospho-(1'-rac-glycerol) (sodium salt) (DPPG$^{(-)}$, $T_m$= 41℃)], and cholesterol were obtained from Avanti Polar Lipids (Alabaster, AL). BODIPY-labelled cholesterol (BODIPY-Chol) and Rhodamine B 1,2-dihexadecanoyl-*sn*-glycero-3-phosphoethanolamine (Rhodamine-DHPE) were purchased from Invitrogen (Carlsbad, CA). Deionized water was obtained from a Millipore Milli-Q purification system. We chose phosphatidylcholine (PC) as a neutral lipid head group and phosphatidylglycerol (PG$^{(-)}$) as a negatively charged lipid head group because the chain melting temperatures of PC and PG$^{(-)}$ lipids with the same acyl tails are almost identical. In cellular membranes, PC is the most common lipid component, and PG$^{(-)}$ is highly representative among charged lipids.

**B. Preparation of giant unilamellar vesicles**

Giant unilamellar vesicles were prepared by the gentle hydration method. Lipids and fluorescent dyes were dissolved in a 2:1(vol/vol) chloroform/methanol solution. The organic solvent was evaporated under a flow of nitrogen gas, and the lipids were further dried under vacuum for 3h. The films were hydrated with 5 μL deionized water at 55 ℃ for 5 min (pre-hydration), and then with 200 μL deionized water or NaCl solution for 1-2 h at 37℃. The final lipid concentration was 0.2 mM. The Rhodamine-DHPE and BODIPY-Chol concentrations were 0.1 μM and 0.2 μM, respectively (0.5% or 1% of the lipid concentration).

**C. Microscopic observation**

The GUV solution was placed on a glass coverslip, which was covered with another smaller coverslip at a spacing of ca. 0.1 mm. We observed the membrane structures with a fluorescent microscope (IX71, Olympus, Japan) and a confocal laser scanning microscope (FV-1000, Olympus, Japan). Rhodamine-DHPE and BODIPY-Chol were used as fluorescent dyes. Rhodamine-DHPE labels the unsaturated lipid-rich phase, whereas BODIPY-Chol labels the cholesterol-rich phase. A standard filter set (U-MWIG) with excitation wavelength $\lambda_{ex}$=530–550nm and emission wavelength $\lambda_{em}$=575 nm was used to monitor the fluorescence of Rhodamine-DHPE, and another set (U-MNIBA) with $\lambda_{ex}$=470–495 nm and $\lambda_{em}$=510-550 nm was used for the BODIPY-Chol dye. The sample temperature was controlled with a microscope stage (type 10021, Japan Hitec).

**D. Definition of membrane morphology**

In this study, we define the membrane morphology in terms of the aperture angle $\theta$, as shown in Fig.1. A spherical shape has $\theta = 0°$, whereas the pore-formed vesicles can have any aperture angle. Based on the aperture angle, we classified pore-formed vesicles as Cup: $\theta < 90°$, Bowl: $90° < \theta < 180°$, and Disc: $\theta = 180°$, as shown in Fig.1.

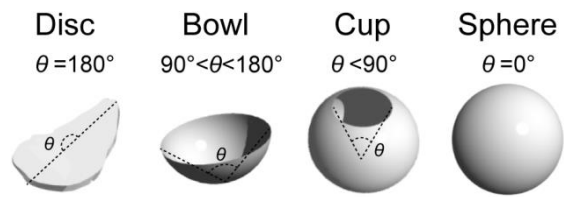

FIG.1. Schematic images of membrane morphologies: disc, bowl, cup, and sphere shapes. Each vesicle can be classified according to the aperture angle as Disc: $\theta = 180°$, Bowl: $90° < \theta < 180°$, Cup: $\theta < 90°$, and Sphere: $\theta = 0°$.

### E. Measurement of morphological transition temperature

In order to reveal the temperature dependence of membrane morphology, we measured the morphological transition temperature between spherical and pore-formed vesicles. We observed the pore-closing process of targeting pore-formed vesicles as increasing in the temperature from room temperature to the desired temperature by 10 ℃/min, and a further delay of 5 min was used in order to approach the equilibrium state. The number of vesicles observed for each composition is n = 50. We then measured the percentage of pore-formed vesicles. The morphological transition temperature is defined as the morphological transition point at which more than 50% of the pore-formed vesicles have disappeared upon heating.

## Ⅲ. RESULTS
### A. Binary mixtures (unsaturated lipid/ saturated lipid)

First, we investigated the effect of a charged lipid on the membrane morphology in charged binary unsaturated/saturated lipid mixtures. In this section, we used the neutral unsaturated lipid DOPC, the neutral saturated lipid DPPC, the negatively charged unsaturated lipid DOPG$^{(-)}$, and the negatively charged saturated lipid DPPG$^{(-)}$. The fluorescent dye Rhodamine-DHPE was used to label the L$_d$ phase, which is the unsaturated lipid-rich phase.

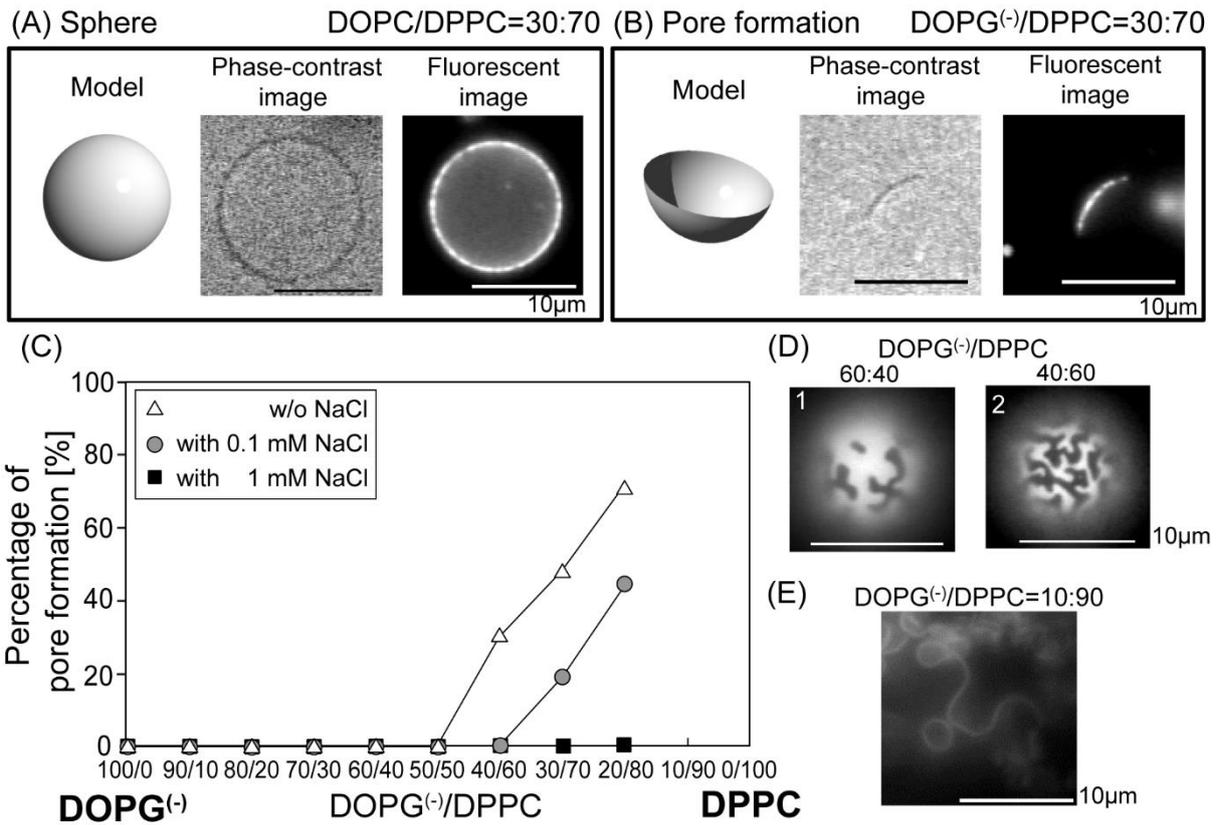

FIG.2. Microscopic images of membrane morphology and percentages of pore formation in a binary mixture of an unsaturated lipid/saturated lipid. Each microscopic image of a GUV is taken at 22°C. (A) Spherical structure for a composition of DOPC/DPPC=30:70 (left: schematic model, center: phase-contrast image, right: fluorescent image). (B) Pore-formed structure of GUV for a composition of DOPG$^{(-)}$/DPPC=30:70. (left: schematic model, center: phase-contrast image, right: fluorescent image). (C) Percentage of pore-formed vesicles in binary mixture of DOPG$^{(-)}$/DPPC at 20°C (open triangle: MQ hydration (w/o NaCl), gray circle: 0.1 mM NaCl, filled square: 1 mM NaCl). Above 90% DPPC, spherical or pore-formed vesicles could not be detected due to the formation of fibrous structures that were connected into a single membrane. (D) Surface structures of GUVs for compositions of DOPG$^{(-)}$/DPPC= 60:40 (image 1), and 40:60 (image 2) w/o NaCl at 20°C. (E) Microscopic image of membrane morphology in binary mixture of DOPG$^{(-)}$/DPPC=10:90. In this composition, string structures which were connected as one membrane were observed.

We observed GUVs with changes in the proportions of the unsaturated and saturated lipids. We observed the membrane morphology using both of phase-contrast and fluorescent microscopy. For neutral lipid mixtures with DOPC/DPPC, vesicles were spherical, as shown in Fig.2A. We replaced the saturated lipid DPPC with DPPG$^{(-)}$, and observed the vesicle shape in DOPC/DPPG$^{(-)}$ mixtures. In this case, the vesicles were also spherical (the microscopic image is not shown). On the other hand, in DOPG$^{(-)}$/DPPC mixtures, we observed bowl-like structure as shown in Fig.2B. The rate of pore formation increased with the DPPC concentration (Fig.2C). At over 90% saturated lipids (DPPC or DPPG$^{(-)}$), we observed string-like structures which were connected as one membrane as shown in Fig.2E, and did not find any spherical or pore-formed vesicles. Therefore, we could not identify the membrane morphology clearly in this compositional region. According to our previous study [18], phase-separated structures were also observed in a binary mixture of DOPG$^{(-)}$/DPPC, as shown in Fig.2D. The white region is the $L_d$ phase, while the black region is the solid-ordered ($S_o$) phase which is a saturated lipid-rich phase.

In addition, we also investigated the membrane morphology in the presence of salt (0.1 mM or 1 mM NaCl). Fig.2C shows the percentages of pore formation in binary mixtures of DOPG$^{(-)}$/DPPC under hydration with Milli Q (MQ) water (open triangles), 0.1 mM NaCl (gray circles), and 1 mM NaCl (filled squares). The number of vesicles observed for each composition is n = 100. When DOPG$^{(-)}$/DPPC vesicles were prepared with 0.1 mM NaCl, the rate of pore formation was decreased compared to that w/o NaCl. In the case of 1 mM NaCl hydration, pore-formed vesicles were not observed. The percentage of pore formation tended to decrease with the salt concentration. These results suggest that the electric charge of the lipid molecule plays an important role in pore formation.

Next, we focused on the relationship between membrane morphology and vesicular size. We measured the membrane diameter of spherical and pore-formed vesicles. In the case of pore-formed vesicles, we used the following method to determine the membrane diameter. First, we measured the membrane surface area of pore-formed vesicles. Second, we calculated the diameter from a measured surface area assuming a spherical shape. We defined this spherical diameter as the membrane diameter of pore-formed vesicles. The percentages of various membrane morphologies for individual membrane diameters in DOPG$^{(-)}$/DPPC=30:70 are summarized in Fig.3. The number of vesicles observed is n=50. Under 2μm, most of the vesicles showed a disc shape. With an increase in diameter, bowl and cup shapes appear. These pore-formed vesicles tend to decrease with an increase in membrane diameter. In addition, over 8 μm, all vesicles are spherical. This result implies that the formation of pore-formed vesicles strongly depends on the vesicular size.

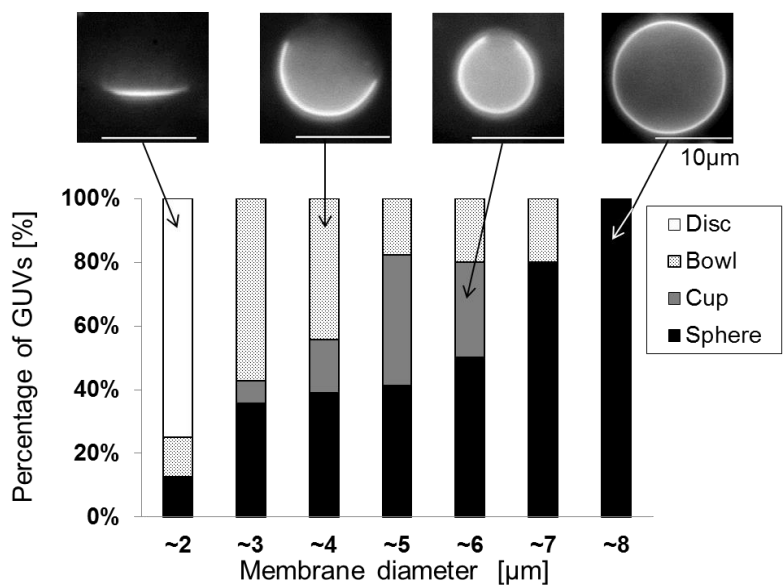

FIG.3. Microscopic images of GUVs and the size-dependence of the fraction of each morphology in a binary mixture of DOPG$^{(-)}$/DPPC=30:70 (white: Disc, light gray: Bowl, gray: Cup, black: Sphere) at 20°C.

Furthermore, we investigated the effect of temperature on the membrane morphology. Fig.4A shows the morphological changes in vesicles with an increase in temperature in DOPG$^{(-)}$/DPPC=30:70. We focused on one of the pore-formed vesicles at 20°C, and increased the temperature to 28°C at 10°C/min. After the temperature reached 28°C, the membrane pore began to close immediately, and a spherical shape finally formed 159 seconds after the temperature reached 28°C. When we cooled the sample solution to 20°C at 20°C/min, this vesicle re-formed a pore (data not shown). We measured this morphological transition temperature according to the procedure mentioned in the Materials and Methods section, and obtained a value of 20°C in DOPG$^{(-)}$/DPPC=30:70. In DOPG$^{(-)}$/DPPC=40:60, pore formation rate was still below 50% even at 20°C (Fig.2C). In our previous study, we investigated the miscibility temperature in a binary mixture of DOPG$^{(-)}$/DPPC [18], and found values of 29℃ and 32℃ in DOPG$^{(-)}$/DPPC=40:60 and 30:70, respectively. Fig.4B shows the percentages of pore formation and the miscibility temperatures at DOPG$^{(-)}$/DPPC=30:70 and 40:60. We found pore-formed vesicles at lower temperatures than the miscibility temperatures. Phase separation may be necessary for the formation of a membrane pore. We will discuss the difference in the morphological change between DOPG$^{(-)}$/DPPC and DOPC/DPPG$^{(-)}$ mixtures, and elaborate the relationship between phase behavior and the membrane morphology in the Discussion.

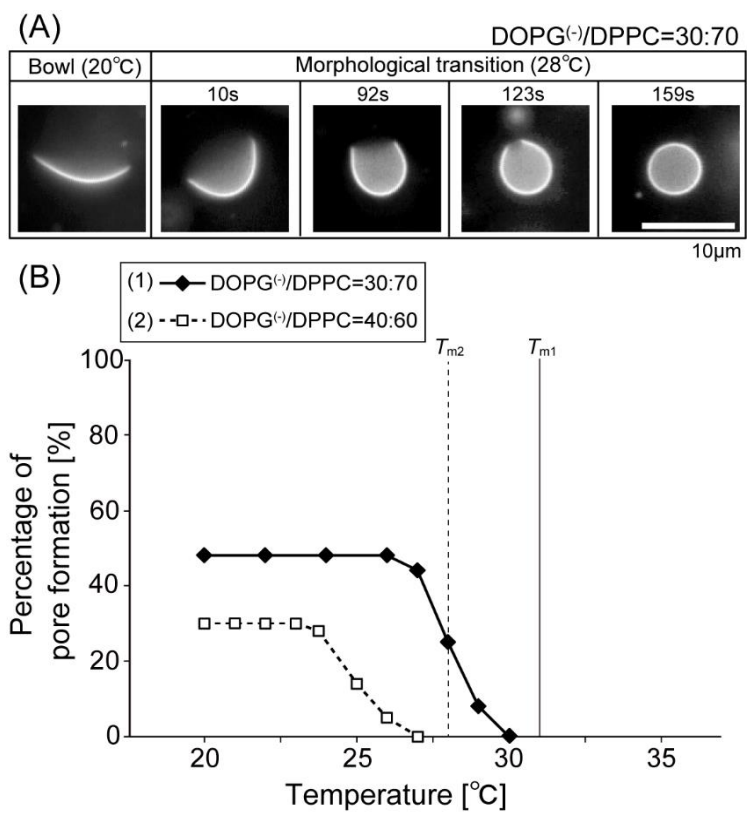

FIG.4. (A) Time course of the morphological change from bowl to sphere in a binary mixture of DOPG$^{(-)}$/DPPC=30:70 with a change in the temperature. We defined the time that the temperature reached 28℃ as 0s. (B) Comparison of the miscibility temperature and morphological transition temperature for DOPG$^{(-)}$/DPPC(filled diamond with solid line: DOPG$^{(-)}$/DPPC=40:60, open square with dashed line: DOPG$^{(-)}$/DPPC=30:70). $T_{m1}$ denoted by vertical solid line and $T_{m2}$ denoted by dashed line show miscibility temperatures in DOPG$^{(-)}$/DPPC=30:70 and 40:60, respectively.

**B. Ternary lipid mixtures (neutral and charged saturated lipids /cholesterol)**

In binary systems composed of an unsaturated lipid and a saturated lipid, pore-formed vesicles were observed only in a negatively charged unsaturated lipid [DOPG$^{(-)}$] and saturated lipid [DPPC] mixture. In addition, the rate of pore formation increased with the DPPC concentration. Therefore, both a charged lipid and a large amount of saturated lipid may be important for observation of pore-formed vesicles. Next, we focused on systems that included neutral and negatively charged saturated lipids. In these saturated lipid mixtures, however, vesicles cannot be formed by the gentle hydration method over the entire compositional range, as mentioned in the previous section. Since the addition of cholesterol induced the phase transition from solid-ordered phase to liquid-ordered phase in the saturated lipid-rich region, where the membrane elasticity is decreased, we added cholesterol to saturated lipid mixtures to form stable vesicles and investigated the effect of charge on the membrane morphology in ternary mixtures composed of neutral and charged saturated lipids and cholesterol.

In this section, we prepared the ternary mixtures of GUVs using the neutral saturated lipid DPPC, the negatively charged saturated lipid DPPG$^{(-)}$, and cholesterol. The fluorescent dyes Rho-DHPE and BODIPY-Chol were used for fluorescent microscope observation. For neutral lipid mixtures of DPPC/Chol=80:20, all of the vesicles assumed a spherical shape. On the other hand, when some fraction of DPPC was replaced with DPPG$^{(-)}$, some vesicles formed pores as described in the previous section. In addition, Fig.5A shows the percentage of pore-formed vesicles in DPPC/DPPG$^{(-)}$/Chol for fixed Chol=20%. The number of vesicles observed for each composition is n = 100. The rate of pore formation increased with the concentration of DPPG$^{(-)}$ (open triangle). At DPPG$^{(-)}$/Chol=80:20, most of the vesicles had pores. We also investigated pore formation in the presence of salt (10 mM NaCl) for a DPPC/DPPG$^{(-)}$/Chol mixture. In Fig.5A, the percentage of pore formation with 10 mM NaCl is indicated by filled triangles. The rate of pore formation was significantly decreased by the addition of salt. As with the binary system, this implies that the negative charge of DPPG$^{(-)}$ plays a critical role in the formation of a membrane pore. As in our previous study [18], in a

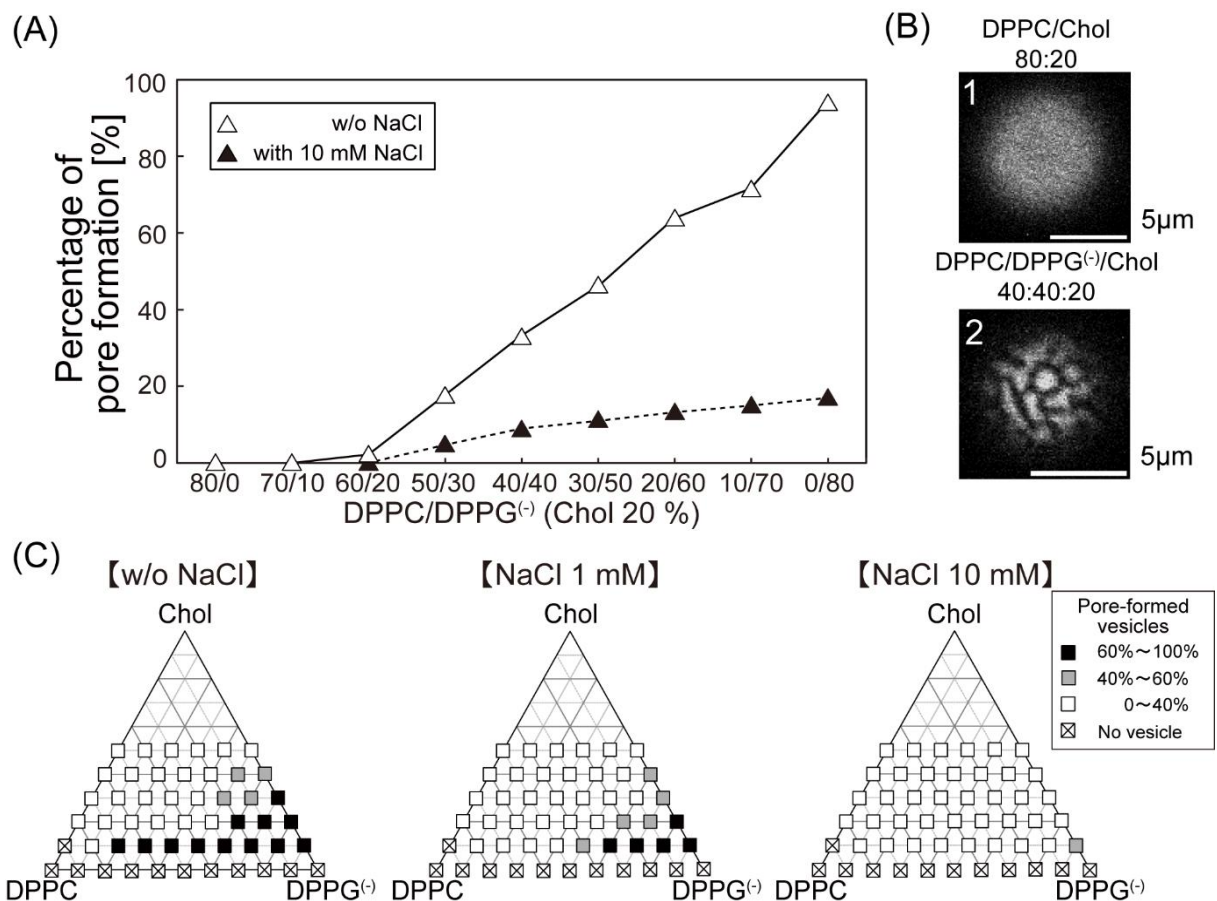

FIG.5. (A) Percentage of pore-formed vesicles in DPPC/DPPG$^{(-)}$/Chol at 20℃ (open triangle: MQ hydration (w/o NaCl), filled triangle: hydration by 10 mM NaCl). (B) Surface structures of GUVs for compositions of DPPC/Chol= 80:20 (image 1), and DPPC/DPPG$^{(-)}$/Chol= 40:40:20 (image 2) w/o NaCl at 20℃. (C) Phase diagrams of DPPC/DPPG$^{(-)}$/Chol mixtures with and without NaCl solutions (left: MQ hydration (w/o NaCl), center: NaCl 1 mM, right: NaCl 10 mM) at room temperature (~22 ℃). Filled, grey, and open squares correspond to systems with 60‑100%, 40‑60%, and 0‑40% pore-formed vesicles, respectively.

DPPC/DPPG$^{(-)}$/Chol mixture, phase-separated structures were observed, as shown in Fig.5B. Since the percentage of the domain area increased with the DPPG$^{(-)}$ concentration, the black region is a DPPG$^{(-)}$-rich S$_o$ phase and the white region is a DPPC and cholesterol-rich L$_o$ phase. The effects of the lipid composition and salt concentration on pore formation are summarized in Fig. 5C. The number of vesicles observed for each composition is over n = 50. The left diagram shows the rate of pore formation for DPPC/DPPG$^{(-)}$/Chol mixtures w/o NaCl. For lower concentrations of DPPG$^{(-)}$, pore-formed vesicles were either not observed or only rarely observed (open squares). On the other hand, the rate of pore formation clearly increases with the DPPG$^{(-)}$ concentration (filled squares).The regions of pore formation with 1 mM and 10 mM NaCl are indicated in the center and right diagrams. As the salt concentration is increased, the rate of pore formation tends to decrease significantly. In particular, most of the vesicles assumed a spherical shape in hydration with 10 mM NaCl. This is because DPPG$^{(-)}$ is screened in the presence of salt and its behavior approaches that of the neutral DPPC.

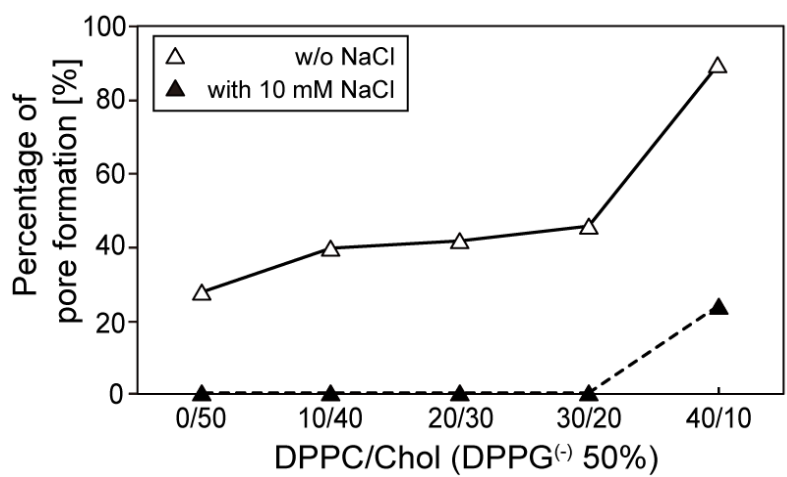

FIG.6. Percentage of pore-formed vesicles in DPPG$^{(-)}$/DPPC/Chol for fixed DPPG$^{(-)}$=50% at 22°C (open triangle: MQ hydration(w/o NaCl), filled triangle: hydration by 10 mM NaCl).

Moreover, pore formation was also affected by the cholesterol concentration. In Fig.6, we changed the ratio between DPPC and cholesterol for fixed DPPG$^{(-)}$=50%, and measured the percentage of pore formation for each composition. The number of vesicles observed for each composition is n = 100. Without NaCl, the rate of pore formation increased in inverse proportion to the cholesterol concentration. In particular, the rate of pore formation dramatically increased between DPPC/DPPG$^{(-)}$/Chol=30:50:20 and 40:50:10. In saturated lipid and cholesterol mixtures, a phase transition can be observed with a change in the cholesterol concentration [38]. At a cholesterol concentration of less than 15%, an $S_o$ phase appears for DPPC. However, when the cholesterol concentration is over 20%, an $L_o$ phase becomes stable. Previous research suggested that the S$_o$ phase has greater bending rigidity than the L$_o$ phase [39]. We expect that the sudden increase of the percentage of pore formation between DPPC/DPPG$^{(-)}$/Chol=30:50:20 and 40:50:10 can be explained by the phase transition from the L$_o$ phase to the S$_o$ phase. This high bending rigidity may play an essential role in pore formation.

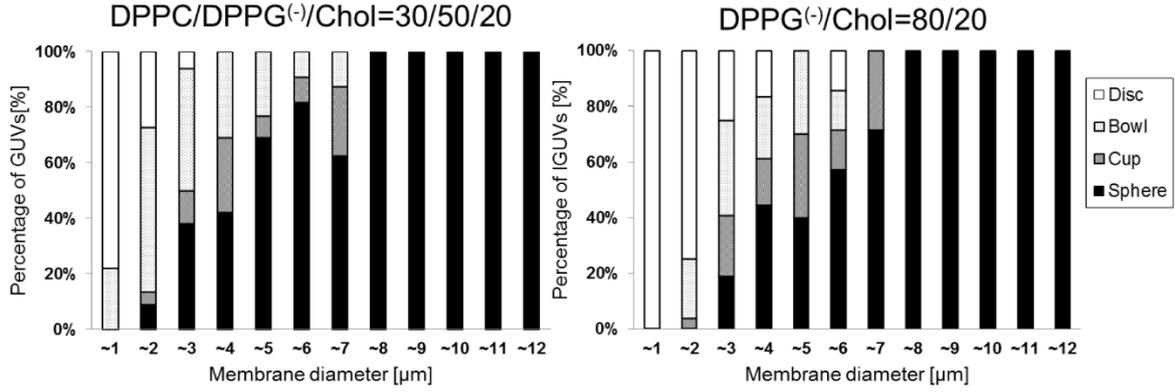

FIG.7. Size distributions of each of the membrane morphologies in DPPC/DPPG$^{(-)}$/Chol=30:50:20, and DPPG$^{(-)}$/Chol =80:20, respectively (white: Disc, light gray: Bowl, gray: Cup, black: Sphere) at 22℃. The number of vesicles observed for each composition is n=50.

The size-dependency of the membrane morphology was examined in the same way as for a binary mixture of DOPG$^{(-)}$/DPPC. We can find a similar size-dependency as in a binary system; the stable shape changes from a disc, to a bowl, cup, or sphere with an increase in the membrane diameter, as shown in Fig. 7. In DPPG$^{(-)}$/Chol=80:20, vesicles tend to form disc and bowl shapes even for a larger diameter compared with DPPC/DPPG$^{(-)}$/Chol=30:50:20.

Furthermore, we observed the temperature-dependence of membrane morphology for pore-formed vesicles. Fig.8A shows morphological changes in vesicles composed of DPPC/DPPG$^{(-)}$/Chol=40:40:20. We focused on vesicles that formed pores at room temperature (~20℃), and increased the temperature to 28℃ at 10℃/min. After the temperature reached 28℃, the membrane morphology immediately changed from a pore-formed structure to a spherical shape. When we reduced the temperature to 20℃ at 20℃/min, vesicles formed pores again (data not shown). Next, we measured the

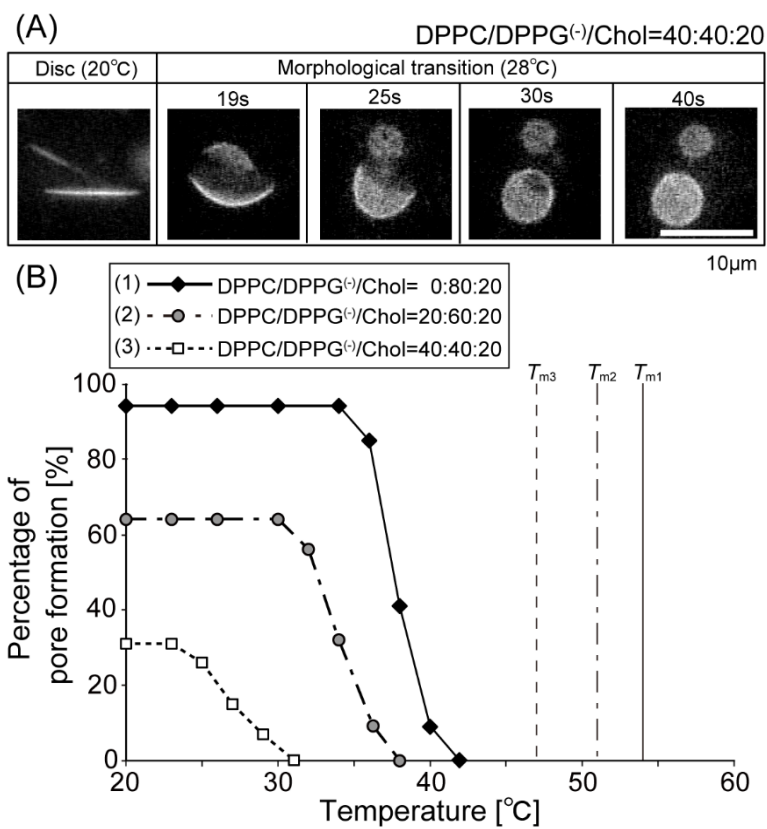

FIG.8. (A) Time course of the morphological change from disc to sphere in a ternary mixture of DPPC/DPPG$^{(-)}$/Chol=40:40:20 with a change in the temperature. We defined the time that the temperature reached 28°C as 0s. (B) Comparison of the miscibility temperature and morphological transition temperature in DPPC/DPPG$^{(-)}$/Chol (filled diamond with solid line: DPPC/DPPG$^{(-)}$/Chol=0:80:20, gray circle with dot-dashed line: DPPC/DPPG$^{(-)}$/Chol=20:60:20, open square with dashed line: DPPC/DPPG$^{(-)}$/Chol=40:40:20). $T_{m1}$ denoted by vertical solid line, $T_{m2}$ denoted by dot-dashed line, and $T_{m3}$ denoted by dashed line show miscibility temperatures in DPPC/DPPG$^{(-)}$/Chol=0:80:20, 20:60:20, and 40:40:20, respectively.

morphological transition temperature for the different compositions of DPPC/DPPG$^{(-)}$/Chol. For DPPC/DPPG$^{(-)}$/Chol=20:60:20 and 0:80:20, the morphological transition temperatures were 33°C and 36°C, respectively. Since pore formation rate did not reach 50% even at 20°C in DPPC/DPPG$^{(-)}$/Chol=40:40:20, we could not obtain the value of the morphological transition temperature. Previously, we investigated the miscibility temperature in a ternary mixture of DPPC/DPPG$^{(-)}$/Chol [18], and observed values of 54°C, 51°C, and 47°C for DPPC/DPPG$^{(-)}$/Chol=0:80:20, 20:60:20 and 40:40:20, respectively. Fig.8B shows the percentages of pore formation and the miscibility temperatures at

DPPC/DPPG$^{(-)}$/Chol=0:80:20, 20:60:20 and 40:40:20. In all cases, the pore-formed vesicles were found at lower temperatures than the miscibility temperatures. There is a larger difference between the morphological transition temperature and the miscibility temperature than in the case of binary mixtures consisting of DOPG$^{(-)}$/DPPC. In addition, the morphological transition temperatures were lower than the phase transition temperature of DPPC and DPPG$^{(-)}$ (41°C).

## Ⅳ. DISCUSSION

According to the experimental results, pore formation can be observed at some lipid compositions, while screening of charged head groups by the addition of salt (NaCl) suppresses pore formation. This suggests that electrostatic interaction plays a key role in the stability of membrane morphology. We discuss the mechanism of pore formation in charged lipid mixtures based on Fromherz's theory [40].

First, we consider the pore-formed vesicles that we observed in the experiments. The elastic free energy of a membrane can be written as

$$E_{(\text{total})} = E_\kappa + E_\gamma \qquad (1)$$

where $E_\kappa$ is the bending energy of the membrane and $E_\gamma$ is the line energy at the edge of a membrane pore. $E_\kappa$ can be obtained from the Helfrich bending energy [41] as

$$E_\kappa = \int \left[ \frac{\kappa}{2}(c_1 + c_2 - c_0)^2 + \kappa' c_1 c_2 \right] dA \qquad (2)$$

where $dA$ is an element of the membrane surface, $c_1$ and $c_2$ are the two principal curvatures, $c_0$ is the spontaneous curvature, $\kappa$ is the bending rigidity, and $\kappa'$ is the Gaussian bending modulus. In addition, $E_\gamma$ can be expressed as

$$E_\gamma = \gamma \int dl \qquad (3)$$

where $dl$ is the line element along the membrane edge, and $\gamma$ is the line tension. The shapes of membranes are determined by competition between the bending and line energies. If the line energy $E_\gamma$ is more dominant than the bending energy $E_\kappa$, the membrane should assume a spherical shape to reduce the line energy. Conversely, if the bending energy $E_\kappa$ becomes dominant, the lipid membrane forms a pore and becomes flat to decrease the bending energy. Previous studies have suggested that charged lipid molecules decrease the line tension $\gamma$ and stabilize the edge of a membrane pore [42,43]. A charged lipid has an apparent bulky head group due to repulsion between the charged head groups and can fit and stabilize the edge of a pore. In addition, counterions released from charged lipids can obtain an increase in entropy in the bending part compared to the flat part [43]. Thus, charged lipids favor the formation of the edge of a pore as do lipids that form membranes with a positive spontaneous curvature [34].

As can be seen in Eq.(2), given a certain shape of a membrane, $E_\kappa$ is determined by the bending rigidity $\kappa$ and the spontaneous curvature $c_0$. Commonly, at room temperature, $\kappa$ of an unsaturated lipid tends to be smaller than that of a saturated lipid. For example, the $\kappa$ of the unsaturated lipid DOPC is $1.5 \times 10^{-20}[J]$ at 21°C [44], while that for the saturated lipid DPPC is $1.0 \times 10^{-18}[J]$ at 23°C [45]. If we assume that the bending rigidity can be characterized by acyl chain structures, the bending rigidities of the charged unsaturated lipid DOPG$^{(-)}$ and the saturated lipid DPPG$^{(-)}$ can be approximated by those of neutral lipids DOPC and DPPC, respectively.

In binary neutral mixtures of DOPC/DPPC, vesicles tend to form spherical shapes, because the increase of line energy at the edge of the membrane by the exposure of hydrocarbon tails to water is prevented. Therefore, $E_\gamma$ has a greater contribution than $E_\kappa$ on a micrometer-length scale. When we replaced the neutral unsaturated lipid DOPC with the charged unsaturated lipid DOPG$^{(-)}$, pore formation was observed in DOPG$^{(-)}$/DPPC=40:60,

30:70, and 20:80. For these compositions, the charged lipid DOPG$^{(-)}$ may be localized at the membrane edge and decrease the line tension $\gamma$. As a result, the bending energy $E_{\kappa}$ becomes more significant than $E_{\gamma}$, and pore-formed structures are observed. At a higher concentration of DPPC such as DOPG$^{(-)}$/DPPC=10:90, we observed the string-like structures. This is because $E_{\kappa}$ is too high to form spherical shapes. At a higher DOPG$^{(-)}$ concentration, $E_{\gamma}$ becomes lower due to reduction of the line tension $\gamma$. However, the bending energy $E_{\kappa}$ is also decreased, since some fraction of DOPG$^{(-)}$ having small bending rigidity $\kappa$ is included in the bulk membrane as opposed to the membrane edge. Thus, the bending energy $E_{\kappa}$ is low enough, and the lipid membrane assumes a spherical shape at a higher DOPG$^{(-)}$ concentration.

On the other hand, when we replaced the saturated lipid DPPC with the negatively saturated lipid DPPG$^{(-)}$, pore formation was not observed for compositions of DOPC/DPPG$^{(-)}$ ranging 10:0 to 20:80. In this case, while the charged lipid DPPG$^{(-)}$ can decrease the line tension, the bulk membrane is composed of the unsaturated lipid DOPC. The bending rigidity $\kappa$ of an unsaturated lipid is 100 times smaller than that of a saturated lipid [44,45]. As a result, since the bending energy $E_{\kappa}$ is low enough, the membrane assumes a spherical shape for all compositions.

We also observed pore-formed structures in neutral and charged saturated lipid/cholesterol mixtures of DPPC/DPPG$^{(-)}$/Chol for fixed Chol=20%. In this mixture, it is believed that DPPC localized at the bulk membrane, whereas DPPG$^{(-)}$ localized at the membrane edge. For cholesterol, we had confirmed in our previous study that it is localized to the DPPC [18]. Thus, the bending energy $E_{\kappa}$ is high for all compositions, since the membrane always includes 80% saturated lipids. In addition, the line energy $E_{\gamma}$ is decreased by the charged lipid DPPG[(-)]. As a result, the percentage of pore-formed structure increases with the DPPG[(-)] concentration (Fig.5). At a lower cholesterol concentration, a saturated lipid-rich region shows an $S_o$ phase [7,46]. With 10% cholesterol, the rate of pore formation was significantly increased (Fig.6). This indicates that $E_\kappa$ increases significantly due to a phase transition to the $S_o$ phase with a decrease in the cholesterol concentration.

In addition, we investigated the size-dependency of the membrane morphology, as shown in Fig.3 and Fig.7. Our experimental results showed that the percentage of pore-formed vesicles decreased with an increase in the membrane diameter. We express the total free energy of Eq. (1) for a pore-formed vesicle with membrane diameter $D$ and aperture angle $\theta$ [28]. Under the assumption that $c_0 = 0$ and $\kappa \sim -\kappa'$ [47], the free energy $F$ of the membrane with a surface area of $\pi D^2$ can be written as,

$$F = E_\kappa + E_\gamma = 2\pi\kappa(1+\cos\theta) + \pi\gamma D \sin\theta \left(\frac{2}{1+\cos\theta}\right)^{1/2}. \tag{4}$$

The first and second terms indicate the bending and line energies, respectively. For simplicity, we compare the energies of a sphere ($\theta = 0$) and disc ($\theta = \pi$). Based on this comparison, the critical line tension $\gamma^*$, which determines the membrane morphology, can be obtained as,

$$\gamma^* = \frac{2\kappa}{D}. \tag{5}$$

When the line tension is larger than the critical line tension, the vesicle assumes a spherical shape. In contrast, a disc shape becomes stable for a line tension smaller than the critical line tension. Interestingly, the critical line tension depends on the membrane diameter, as shown in Eq. (5). For a larger membrane diameter, the line tension easily becomes larger than the critical line tension, since the critical line tension is decreased. Therefore, a spherical vesicle is often seen for a larger membrane diameter. As shown in Fig.3 and Fig.7, the stable vesicle shape changes at around $D \sim 5\,\mu\text{m}$. Using $D \sim 5\,\mu\text{m}$, we estimate the two limiting values of the critical line tension. In one case, the bulk membrane consists solely of a saturated lipid and we use $\kappa = 1.0 \times 10^{-18} [J]$. The critical line tension can be estimated as $\gamma^* = 0.4 \, \text{pN}$. In the other case, the bulk membrane consists only of an unsaturated lipid and we choose $\kappa = 1.5 \times 10^{-20} [J]$. In this case, the critical line tension becomes $\gamma^* = 6.0 \times 10^{-3} \, \text{pN}$. In our experiments, a pore-formed vesicle can be expected to have a line tension between these two limiting values. It has been reported that the line tension at the edge of a pore is about 10 pN [48]. Therefore, charged lipids can contribute to a two or three orders of magnitude reduction in line tension. If we wish to understand the mechanism of pore stability produced by charged lipids, it will be important to measure the line tension quantitatively in the future.

In both binary and ternary systems, pore formation is suppressed by the addition of salt. Due to screening of the repulsion between charged head groups, the apparent size of a charged head group becomes smaller. In other words, the spontaneous curvature of membranes consisting of charged lipids effectively becomes smaller, and charged lipids prefer to form a planar membrane. Therefore, since charged lipids cannot stabilize a highly curved edge region, it is difficult to find pore-formed vesicles in the presence of salt.

According to our hypothesis, charged lipid molecules are localized at the edge of a membrane pore, while a large amount of neutral lipid molecules is included the bulk membrane. This structure can be regarded as phase separation. As we have reported previously [18], phase separation often occurs in this lipid mixture. Furthermore, as the temperature increased, the pore-formed vesicles disappeared before the miscibility transition temperature, as shown in Fig.4 and 8. Thus, it is possible that a phase-separated structure triggers pore formation. Although it is difficult to detect the localization of negatively charged lipids at the edge of a pore by using microscopic observation due to the lack of suitable fluorescent probes for labeling a DPPG$^{(-)}$-rich phase, we intend to investigate this point in detail in the future.

In our experiments, GUVs were prepared by the gentle hydration method, as mentioned in the Materials and Methods section. This swelling method was known to produce various different shapes such as elliptical, oblate, tubular, and stomatocytic vesicles. Although some non-spherical vesicles were observed in neutral lipid mixtures (DOPC/DPPC and DPPC/Chol), we could not find any pore-formed vesicles, in other words, all observed vesicles were closed. In addition, non-spherical vesicles were hardly observed in charged lipid mixtures (DOPG$^{(-)}$/DPPC and DPPC/DPPG$^{(-)}$/Chol). Moreover, all pore-formed vesicles changed to spherical shape by increasing temperature as shown in Fig.4 and Fig.8. Therefore, we believe that the non-spherical shape was not an important factor that affects the pore formation, and focus only on a spherical shape as a closed vesicle to observe and discuss the membrane morphology.

Finally, we reproduced the results of an experimental system using coarse-grained molecular dynamics simulations. It is difficult to observe the structure at the edge of pore and the dynamics of pore formation on the molecular scale experimentally. Therefore, we performed the coarse-grained molecular dynamics simulation to understand the mechanism of pore formation in detail. We extended the model described by Cooke and coworkers [49] to explain the behavior observed in charged lipid membranes. For simplicity, we calculated binary charged lipid membranes. A lipid molecule is represented by one hydrophilic bead and two hydrophobic beads, and these beads are connected by bonds. The excluded-volume interaction between two beads separated by a distance $r$ is

$$V_{\text{rep}}(r;b) = \begin{cases} 4v\left[\left(\dfrac{b}{r}\right)^{12} - \left(\dfrac{b}{r}\right)^{6} + \dfrac{1}{4}\right], & r \leq r_c \\ 0 & r > r_c \end{cases} \qquad (6)$$

where $r_c = 2^{1/6}b$. $v$ and $\sigma$ are the units of energy and length, respectively. We choose $b_{\text{head,head}} = b_{\text{head,tail}} = 0.95\sigma$ and $b_{\text{tail,tail}} = \sigma$ to form a stable bilayer. The potentials for the stretching and bending of bonds between connected beads are described as

$$V_{\text{bond}}(r) = \frac{1}{2} k_{\text{bond}} (r - \sigma)^2, \tag{7}$$

and

$$V_{\text{bend}}(\psi) = \frac{1}{2} k_{\text{bend}} (1 - \cos\psi)^2, \tag{8}$$

where $k_{\text{bond}} = 500v$ is the bond stiffness, $k_{\text{bend}} = 60v$ is the bending stiffness, and $\psi$ is the angle between adjacent bond vectors. The attractive potential between hydrophobic beads is expressed as

$$V_{\text{attr}}(r) = \begin{cases} -v & r < r_c \\ -v \cos^2\left[\dfrac{\pi(r - r_c)}{2w_c}\right], & r_c \leq r \leq r_c + w_c \\ 0 & r > r_c + w_c \end{cases} \tag{9}$$

The cutoff length for this attractive potential is $w_c$. When $w_c$ is large, the lipid is in a solid phase. In contrast, the lipid is in a liquid phase when $w_c$ is small [49]. Finally, the electrostatic repulsion between charged head groups is expressed as the Debye-Hückel potential

$$V_{\text{elec}}(r) = v \ell_B z_1 z_2 \frac{\exp(-r/\ell_D)}{r}, \tag{10}$$

where $\ell_B = \sigma$ is the Bjerrum length, $z_1$ and $z_2$ are the valencies of charged head groups, and $\ell_D = \sigma\sqrt{(\varepsilon k_B T)/(n_0 e^2)}$ is the Debye-Hückel screening length which is related to the bulk salt concentration $n_0$ ($\varepsilon$, $k_B$, $T$, and $e$ are the dielectric constant of the solution, Boltzmann constant, temperature, and elementary charge, respectively). Since the negatively charged head group PG$^{(-)}$ has a monovalent ion, we set $z_1 = z_2 = -1$. Moreover, we did not set the cut-off length for the electrostatic interaction.

The lipids obey the stochastic dynamics described by the Langevin equation

$$m \frac{d^2 \mathbf{r}_i}{dt^2} = -\eta \frac{d\mathbf{r}_i}{dt} + \mathbf{f}_i^U + \boldsymbol{\xi}_i, \tag{11}$$

where $m=1$, and $\eta=1$ are mass and drag coefficients, respectively. The force $\mathbf{f}$ is obtained from Eqs.(6)~(10). The constant $\tau=\eta\sigma^2/v$ is chosen as the unit for the time scale, and the time step is set at $dt=7.5\times10^{-3}\tau$. The Brownian force $\xi_i$ satisfies the fluctuation dissipation theorem

$$\langle \xi_i(t)\xi_i(t')\rangle = 6v\eta\delta_{ij}\delta(t-t'). \tag{12}$$

We performed calculations for two binary systems consisting of neutral and negatively charged lipids that correspond to DOPC/DPPG$^{(-)}$ and DOPG$^{(-)}$/DPPC. The total number of lipid molecules is 5000, and the mixing ratio is unsaturated lipid: saturated lipid=50:50. For $v=1$, we choose $w_{c(\text{DOPC–DOPC})}=1.6\sigma$, $w_{c(\text{DPPG–DPPG})}=1.8\sigma$, and $w_{c(\text{DOPC–DPPG})}=1.5\sigma$ for DOPC/DPPG$^{(-)}$ and $w_{c(\text{DOPG–DOPG})}=1.6\sigma$, $w_{c(\text{DPPC–DPPC})}=1.8\sigma$, and $w_{c(\text{DOPG–DPPC})}=1.5\sigma$ for DOPG$^{(-)}$/DPPC. In a neutral system, a lipid is in a liquid phase for $w_c=1.6\sigma$ and in a solid phase for $w_c=1.8\sigma$ [49]. The lipids are randomly set in a bilayer membrane at the initial state as shown in Fig.10A.

We show the further information about the numerical simulation. We performed the calculations until $t=75000\tau$, and confirmed that the free energy did not changed significantly around $t=75000\tau$. Since the membrane bending rigidity $\kappa$ equals 8-30$k_\text{B}T$ in the range of $w_c=1.5-1.8\sigma$ for neutral lipids [49], these cut-off lengths which we used are realistic values. The line tension is determined by the balance among $w_{c(\text{S-S})}$, $w_{c(\text{U-U})}$, and $w_{c(\text{S-U})}$, where the subscripts S and U mean saturated and unsaturated lipids, respectively. Although it is difficult to obtain the actual value of the line tension explicitly, the phase-separated domains are budded for neutral systems at $w_{c(\text{S-S})}=w_{c(\text{U-U})}=1.5\sigma$ and $w_{c(\text{S-U})}=1.3\sigma$ [49]. Under the values in our simulations, the domain-induced budding does not occur, and we regard these values as reasonable. We did not consider the water molecules explicitly. Instead, they are taken into account in the attraction between hydrophobic beads in Eq. (9) and the electrostatic interaction in Eq. (10) implicitly.

| | DOPC/DPPG$^{(-)}$ | DOPG$^{(-)}$/DPPC |
|---|---|---|
| Salt concentration 1 M | (A) 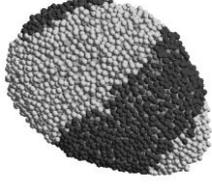 | (B) 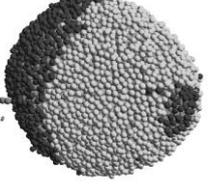 |
| Salt concentration 100 mM | (C) 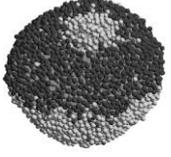 | (D) 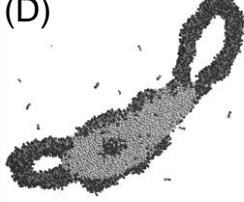 |

FIG.9. Coarse-grained molecular dynamics simulations of binary mixtures consisting of unsaturated lipid and saturated lipids. Dark and bright beads represent charged and neutral lipids, respectively. (A) DOPC/DPPG$^{(-)}$ in 1 M salt solution, (B) DOPG$^{(-)}$/DPPC in 1 M salt solution, (C) DOPC/DPPG$^{(-)}$ in 100 mM salt solution, (D) DOPC/DPPG$^{(-)}$ in 100 mM salt solution. $t = 75000\tau$.

We calculated four simulations from different initial conditions, and confirmed the repeatability. Typical snapshots of the coarse-grained MD simulation are shown in Fig.9. When the salt concentration in the solution is high (1 M), the vesicle shape is spherical in both DOPC/DPPG$^{(-)}$ and DOPG$^{(-)}$/DPPC systems, as shown in Fig.9A and B. In addition, we can see phase-separated structures. Although we found one pore-formed vesicle out of four simulations, the vesicle shape of DOPC/DPPG$^{(-)}$ under a low salt concentration (100 mM) is practically spherical (Fig.9C). A disc shape is always observed for DOPG$^{(-)}$/DPPC, as shown in Fig.9D. In the experiment, since we observed pore formation in DOPG$^{(-)}$/DPPC as shown in Fig.2, and it was difficult to find any morphological changes in DOPC/DPPG$^{(-)}$, these experimental results are consistent with the results from numerical simulations as shown in Fig.9C and D. In addition, the pore formation is suppressed by adding salt in the experiment as shown in Fig.2C. The results from the numerical simulations also indicate that the pore formation (transition from spherical to disc shape) could not be found at higher salt concentration as shown in Fig.9B and D. Our numerical simulations gave results consistent with those from the experiments. However, the orders of salt concentration are different between experiments and numerical simulations. In numerical simulations, the vesicle size is much smaller than the micrometer-sized vesicle observed in experiment. The critical line tension expressed by Eq. (5) can be larger in the case of the vesicle in numerical simulation due to the smaller diameter. Assuming the diameter of the order of 10 nm, the critical line tension is of the order of 1 - 100 pN, which is comparable to the reported experimental value of 10 pN [46]. Namely, based on the above discussion, the small charged vesicles in numerical simulation can form pores even in higher salt concentration than the larger ones in the experiment. Although there are quantitative differences between experiment and simulation, we can see the qualitative same salt concentration-dependence of the morphological change. We believe that this result from numerical simulations supports the experimental observations.

Interestingly, the charged lipids that are shown in dark are localized at the edge of the disc in numerical simulations as shown in Fig.9D. Since the electrostatic repulsion between charged head groups starts to increase, the charged lipids with the apparent bulky head group are localized at the edge of the disc to reduce the line energy, rather than close the pore. This result supports our experimental findings and theoretical predictions.

The typical time course of pore formation is shown in Fig.10. The time for pore appearance is about $t = 2512 \pm 537\tau$. A pore appeared in the charged lipid-rich region, and this pore became larger as time progressed. As phase separation progressed, another pore appeared in the charged lipid-rich region. Finally, the pores became larger and the vesicle shape approached a disc as shown in Fig.9D.

We mention the mechanism for pore-opening based on the results from numerical simulations. The pore appears at the charged lipid-rich region as shown in Fig.10. Therefore, the charged lipid-rich region stored the large electrostatic tension as domain coarsening. When the repulsion between charged head groups overcomes the attraction between hydrophobic tails, a pore is formed in the charged lipid-rich region to eliminate this tension. Coarse-grained Monte Carlo simulations reported pore formation in charged lipid systems [50]. In this study, pore formation was observed when the repulsion between charged head groups was relatively large, and this result is consistent with our results. Therefore, charged lipids play an important role in the pore-opening process as well as in stabilization of the pore edge.

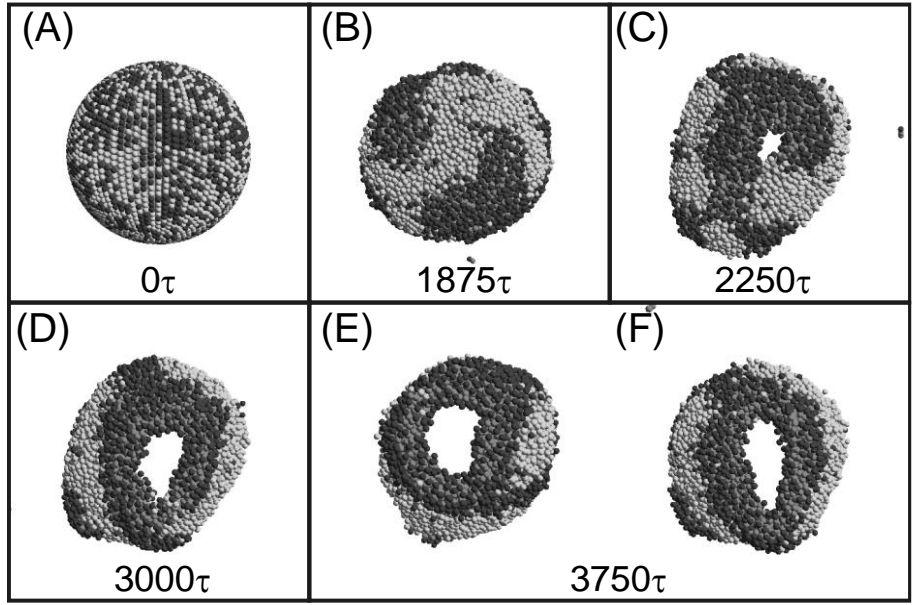

FIG.10. Numerical simulation of pore formation dynamics for DOPG$^{(-)}$/DPPC = 50:50 in 100 mM salt solution, which corresponds to the condition in Fig.9D. Dark and bright beads represent charged (DOPG$^{(-)}$) and neutral (DPPC) lipids, respectively. Each image represents one surface of vesicles. (E) and (F) show other surfaces. (A) $t = 0\tau$, (B) $t = 1875\tau$, (C) $t = 2250\tau$, (D) $t = 3000\tau$, (E) and (F) $t = 3750\tau$. After phase separation progressed prior to pore formation in (B), a pore appeared in the charged lipid-rich region, as shown in (C), and the pore became larger as time progressed, as shown in (D) and (E). As phase separation progressed, another pore appeared in the charged lipid-rich region in (F).

In Fig. 4 and 8, the temperature dependence of membrane morphology was shown. Even if phase-separated structures are observed, pore formation cannot be found at higher temperature. We discuss why the pore was closed at a higher temperature still below the miscibility temperature. On approaching the miscibility temperature, the lipid concentration difference between two separated phases becomes smaller due to the mixing entropy. This means that the neutral lipid concentration in the charged lipid-rich region becomes higher. As shown in Fig.9D, there is a cap structure consisting mainly of charged lipids at the edge of a pore. This cap can be regarded as a highly-curved monolayer. If the neutral lipid concentration in this cap region becomes higher, the spontaneous curvature of the monolayer is decreased. This reduction in the spontaneous curvature leads to an increase in the line tension at the pore edge. To avoid the line energy loss, the pore will be closed. Furthermore, the bending energy may be an important factor. On increasing in the temperature, it was reported that the bending rigidities of two phases (the $L_o$ and $L_d$ phases) become smaller and they approach each other in DOPC/DPPC/Chol ternary mixtures [51]. The phase separation in our experiments occurred between $S_o$ and $L_d$ phases in the binary mixtures, and $S_o$ and $L_o$ phases in the ternary mixtures. Although the phase-separated phases are different between our experiment and the previous work, we can speculate that the bending rigidities of two separated phases may become smaller on increasing the temperature in our systems. If the bending rigidities are decreased, the bending energy is also decreased. In particular, the bending energy in the bulk membrane may be significantly decreased, since the membrane area of the bulk membrane is much larger than that of the edge region. Therefore, there is no large bending energy loss to form the spherical shape. In future, it is important to investigate the temperature dependences of the bending rigidity of each phase and the line tension to understand the pore-closing behavior more precisely.

## Ⅴ. CONCLUSION

In this work, we investigated the effect of charge on the membrane morphology and discussed the relationship between membrane phase separation and the membrane morphology. We observed the membrane morphology in various mixtures containing charged lipids. In binary mixtures of unsaturated lipid/saturated lipid, a neutral composition of DOPC/DPPC formed a spherical shape, whereas pore-formed structures were observed in DOPG$^{(-)}$/DPPC mixtures. Pore formation was suppressed by the addition of salt due to the screening of the electric charge of DOPG$^{(-)}$. In ternary mixtures consisting of DPPC/DPPG$^{(-)}$/Chol, pore-formed structures were also observed. The rate of pore formation increased with the DPPG$^{(-)}$ concentration. In the presence of salt, pore formation was suppressed as with a binary mixture. These results revealed that electrostatic interaction between charged lipids and sodium ions can change the elastic energy of a lipid membrane and result in morphological changes such as pore-opening and -closing. Moreover, the results from an experiment and a theoretical model corresponded to the results of coarse-grained molecular dynamics simulations.

Our study showed that electrostatic interaction between charged lipids and a cation in solution significantly affected phase behavior [18] and the membrane morphology. In our experiments, the salt concentrations were 0.1, 1, and 10 mM, respectively. These concentrations were lower than the concentration in physiological conditions of living cells, where salt concentration is about ~150 mM. However, our experimental results showed morphological changes on a micrometer scale, whereas morphological changes of living biomembranes occur on a nanometer scale. In our numerical simulation, a membrane pore was formed in even 150mM. Therefore, it is possible that morphological control of the nanometer scale membrane needs a high concentration of salt as in vivo.

In living cells, which are more complicated than model systems, membrane morphology may be controlled by a change of salt concentration such as calcium signaling [52]. In addition, it is worth pointing out that charged lipids may be localized at the edge of the membrane pore and stabilize the membrane pore. These effects should contribute to the control of membrane morphological changes in living cells. Our findings should contribute to an understanding of the mechanism of dynamic processes in biomembranes, including endocytosis, autophagy, and vesicular transport.

**ACKNOWLEDGMENTS**

Technical assistance from Ms Ryoko Sugimoto and Mr Masato Amino is greatly appreciated. We thank Dr S. Komura, Dr D. Andelman, Dr M. Hishida, Dr W. Shinoda and Dr T. Taniguchi for their fruitful discussions. This work was supported by a Sasakawa Scientific Research Grant from The Japan Science Society, by MEXT KAKENHI (Nos. 25104510, 26103516, 26800222, 15H00806, 13J01297), and by the Kao Foundation for Arts and Sciences, and by the Kurata Memorial Hitachi Science and Technology Foundation.